\documentclass[aps,prc,twocolumn,showpacs,groupedaddress,nofootinbib]{revtex4}  
\usepackage{graphicx}  
\usepackage{dcolumn}  
\usepackage{bm}        
\usepackage{amssymb}

\begin{document}

\title{Evidence of delayed light emission of TetraPhenyl Butadiene excited by liquid Argon scintillation light}

\author{E.~Segreto}
\email[email: ]{ettore.segreto@lngs.infn.it, segreto@ifi.unicamp.br}
\affiliation{Instituto de Fisica ''Gleb Wataghin'' Universidade Estadual de Campinas - UNICAMP - Rua Sergio Buarque de Holanda, No 777, CEP 13083-859 Campinas, Sao Paulo, Brazil}


\begin{abstract}
TetraPhenyl Butadiene is the wavelength shifter most widely used in combination with liquid Argon. The latter  emits scintillation photons with a wavelength of 127 nm that need
to be downshifted to be detected by photomultipliers with glass or quartz windows. TetraPhenyl Butadiene has been demonstrated to have an extremely high conversion efficiency, 
possibly higher than 100\% for 127 nm photons, while there is no precise information about the time dependence of its emission. It is usually assumed to be exponentially decaying 
with a characteristic time of the order of one ns, as an extrapolation from measurements with exciting radiation in the near UV. 
This work shows that TetraPhenyl Butadiene, when excited by 127 nm photons, reemits photons not only with a very short decay time, but also with slower ones due to 
triplet states de-excitations. This fact can strongly contribute to clarify the anomalies in liquid Argon scintillation light reported in literature since seventies, namely the inconsistency 
in the measured values of the long decay time constant and the appearance of an intermediate component.  
Similar effects should be also expected when the TPB is used in combination with Helium and Neon, that emit scintillation photons with wavelengths shorter than 127 nm.  
\end{abstract}

\pacs{29.40.Mc,33.50.Dq,33.80.Eh,61.25.Bi,78.47.jd}
\maketitle

\section{\label{sec:introduction}Introduction}
Liquid Argon (LAr) is  a widely used active medium in particle detectors, especially in the fields of neutrino physics and Dark Matter direct search 
\cite{warp_100,icarus,darkside,micro}. It is often used in scintillation detectors thanks to its high photon yield ($\sim$ 40000 photons/MeV at null electric field for m.i.p.) and to the
possibility of discriminating different ionizing particles through pulse shape discrimination techniques (see for instance \cite{phd_rob}). The wavelength of the emitted radiation is 
around 127 nm, so in the vacuum UV (VUV). The most efficient and viable way of detecting LAr scintillation light is to downshift it to longer wavelengths, where common quartz or 
glass windowed photo-devices are sensitive.\\
 The most popular wavelength shifter used in combination with LAr is TetraPhenyl Butadiene (TPB) \cite{tpb_paper,lally,mckinsey}, that has been shown to have an extremely high 
 efficiency in converting VUV photons into visible ones (possibly higher than 100$\%$  \cite{gehman}). \\
 On the other side there isn't a precise knowledge of the TPB emission time spectrum when excited by 127 nm photons.
  It is usually described by a single decaying exponential with characteristic time  in the range of 1 ns, as an extrapolation from measurements performed with exciting radiation in 
  the range of the near UV (around 350 nm) \cite{flournoy,camposeo}.\\
  This is perfectly compatible with the photo-excitation of singlet states (S$_n$) of the $\Pi$ electrons of the TPB molecules. They decay via internal conversion to the first 
  excited  singlet state S$_1$  in less than one ns.  The scintillation photon is produced by the radiative de-excitation of this state to the fundamental state (S$_1$ $\rightarrow$  
  S$_0$) that typically has a characteristic time of the order of one ns \cite{laustriat,birks,birks2}.\\
   The point never considered up to now is that VUV scintillation photons' energy (9.7 eV) could very likely exceed the ionization potential of TPB. 
  Actually  there are no available data in the literature, but a calculation leads to a value of 5.4 eV \cite{kafer}. This could appear an extremely low energy, but it is worth 
 noticing that similar compounds like p-terphenyl and anthracene, both used as scintillators or wavelength-shifters, have ionization energies between 7 and 8 eV, not so far 
 from that estimated for TPB. Furthermore it is not difficult to find examples of conjugated molecules with response similar to that of TPB in the UV-vis region with ionization
  energies in the range of 5-6 eV, as PTCDA, Alq3 or CuPc \cite{hill}.\\
 TPB molecules are very likely ionized by LAr scintillation photons and the emitted electron would have enough energy to excite singlet or triplet states of some of the surrounding 
  molecules. Also the recombination of the electron-ion pair can lead to the population of triplet states. Excited singlet states produce the so called prompt fraction of 
  scintillation within few ns through the de-excitation of the S$_1$ states to the ground level. Excited triplet states decay very fast to the lowest lying triplet state T$_1$ via internal 
  conversion. These long lived states (the transition T$_1$ $\rightarrow$ S$_0$ is forbidden by selection rules) are the precursors of the delayed fraction of the scintillation  in pure 
  aromatic media through the  triplet-triplet interaction process:  T$_1$ + T$_1$ $\rightarrow$ S$_1$ + S$_0$, where the scintillation photon is produced by the de-excitation of the 
  S$_1$ state \cite{laustriat}. \\
In this paper experimental evidence of the existence of a delayed component of the scintillation light of TPB excited by LAr VUV scintillation photons is presented.
 A similar effect has already been  reported, for example, for sodium salycilate and for p-terphenyl (\cite{baker,klein} and references therein).\\
This experimental fact can clarify some of the {\emph{anomalies}} of LAr scintillation reported in literature. Namely
the inconsistency in the measured values of the long decay time constant and the 
appearance of an unexpected intermediate component between the fast and the slow ones  \cite{nitrogen,lippincott,heindl,morikawa}.
TPB ionization easily explains the observation of conversion efficiencies for  LAr VUV photons higher that 100$\%$, since each absorbed photon could excite more than one 
TPB molecule at a time.\\

The same effect must also be present when TPB is used to downshift the scintillation light of liquid Neon and of liquid Helium,  that are more energetic than LAr 
ones, since they have wavelengths around 80 nm. In particular it could have a role in explaining some of the not fully understood features in their time dependence.  

\section{\label{sec:iexp_meth_and_setup}Experimental evidence}
Measuring the time response of TPB at 127 nm is an extremely difficult task, because a fast pulsed 127 nm light source ($\sim$ few ns FWHM)  is 
required, together with a  system that allows to drive VUV photons on a TPB  layer.  The experimental approach used in this work is based on the features of the LAr scintillation 
light itself 
and in particular profits of  the fact that it can be reduced to a very fast pulse if the liquid is heavily contaminated by Nitrogen. 
The effects of Nitrogen contaminations on LAr scintillation light have been extensively studied in \cite{nitrogen} and the time response of TPB to 127 nm photons can be directly 
taken from there. In that work, in fact, the scintillation light, quenched by 
N$_2$, is wave-shifted by a TPB layer and then detected by a photomultiplier and at any level of N$_2$ contamination studied, the probability density function in time  (p.d.f.) of the 
photons has been measured.  The p.d.f. at the highest concentration (3000 ppm) can be confidently interpreted as the time response of TPB to 127 nm photons.  
At that concentration, LAr scintillation is reduced to a pulse of the 
duration of few ns and consequently all the features observed in the p.d.f of the detected photons  must be attributed to the time dependence of TPB fluorescence.  It will be shown 
in section \ref{sec:vuv_lar_photons} that it has a non trivial shape and that it contains a delayed component together with the expected prompt/instantaneous one.\\
 
In order to check that the p.d.f. measured with LAr VUV photons effectively represents the response of TPB and is not due to any side effect or uncontrolled systematics, like the
 unwanted  pollution of the liquid by unknown contaminants, a dedicated experimental test has been performed. A TPB film has been directly irradiated with $\alpha$ and $\beta$ 
 particles and the p.d.f. of its scintillation light has been measured and compared with the one obtained with VUV photons in LAr.\\ 
  The comparison allows to demonstrate that the delayed scintillation in the time response of the TPB to 127 photons is genuine and is a consequence of the triplet-triplet interaction 
  process.
  
 \subsection{TPB response to LAr scintillation photons}
 \label{sec:vuv_lar_photons}
The scintillation light 
of LAr proceeds through the de-excitation of the excited dimer Ar$_2^*$ and shows two decay components: one very fast ($\sim$ 6 ns) originating from the decay of the 
lowest-lying  singlet state - $^{1}\Sigma$ - and one very slow ($\sim$ 1.3 $\mu$s) from the decay of the lowest-lying triplet state - $^{3}\Sigma$ - \cite{kubota,doke}. Sometimes an 
intermediate  component with decay time of the order of 100 ns has been observed by experimental groups  \cite{nitrogen,lippincott,morikawa}, not expected on the basis of the 
accepted theory of LAr scintillation mechanism.

It has been clearly shown \cite{nitrogen,himi} that N$_2$ contaminations in LAr produce a quenching of the scintillation light, while no other emission phenomenon from N$_2$ 
has  been observed even at extremely high levels of contaminations ($\sim$ 10$\%$).
The quenching process is a collisional one and the net effect is that the decay times of LAr scintillation components are shortened according to:
\begin{equation} 
\frac{1}{{\tau}'_{f,s}([N_2])} = \frac{1}{\tau_{f,s}} + k_q\times [N_2]
\end{equation}
and consequently the relative abundances of the fast and slow components become:
\begin{equation}
A'_{f,s}([N_2]) = \frac{A_{f,s}}{1+\tau_{f,s}\times k_q \times [N_2]}
\end{equation}

where $\tau_{f,s}$ and $A_{f,s}$ are the decay times and amplitudes of the fast/slow component for uncontaminated LAr \footnote[1]{It is assumed here that the probability density
 function for scintillation photons is $A_f/\tau_f~exp(-t/\tau_f) + A_s/\tau_s~exp(-t/\tau_s)$ and  $A_f + A_s = 1 $}, $[N_2]$ is the nitrogen contamination in ppm and  $k_q$ is the
  reaction rate that has been measured to be $k_q = 0.11\pm 0.01~{\mu}s^{-1}ppm^{-1}$  \cite{nitrogen}.
      
  Taking into account that for $\gamma$/e$^-$ excitations of uncontaminated LAr, $A_f = 0.25$, $A_s=0.75$, it can be easily found that for $[N_2]$ = 3000 ppm one obtains 
  ${\tau}'_f \simeq$ 2 ns, ${\tau_s}'\simeq$ 3 ns, $A'_f$ =  0.1 and $A'_s $= 1.4 $\times$ 10$^{-3}$. In general any additional physical scintillation component of LAr would have a 
  decay time below 3 ns.\\ 
   The scintillation light in heavily N$_2$ doped LAr  results to be a very fast  pulse that is ideal to study the TPB response to 127 nm photons.\\
  
\begin{figure}[h]
\includegraphics*[width=8.6cm,height=8.6cm]{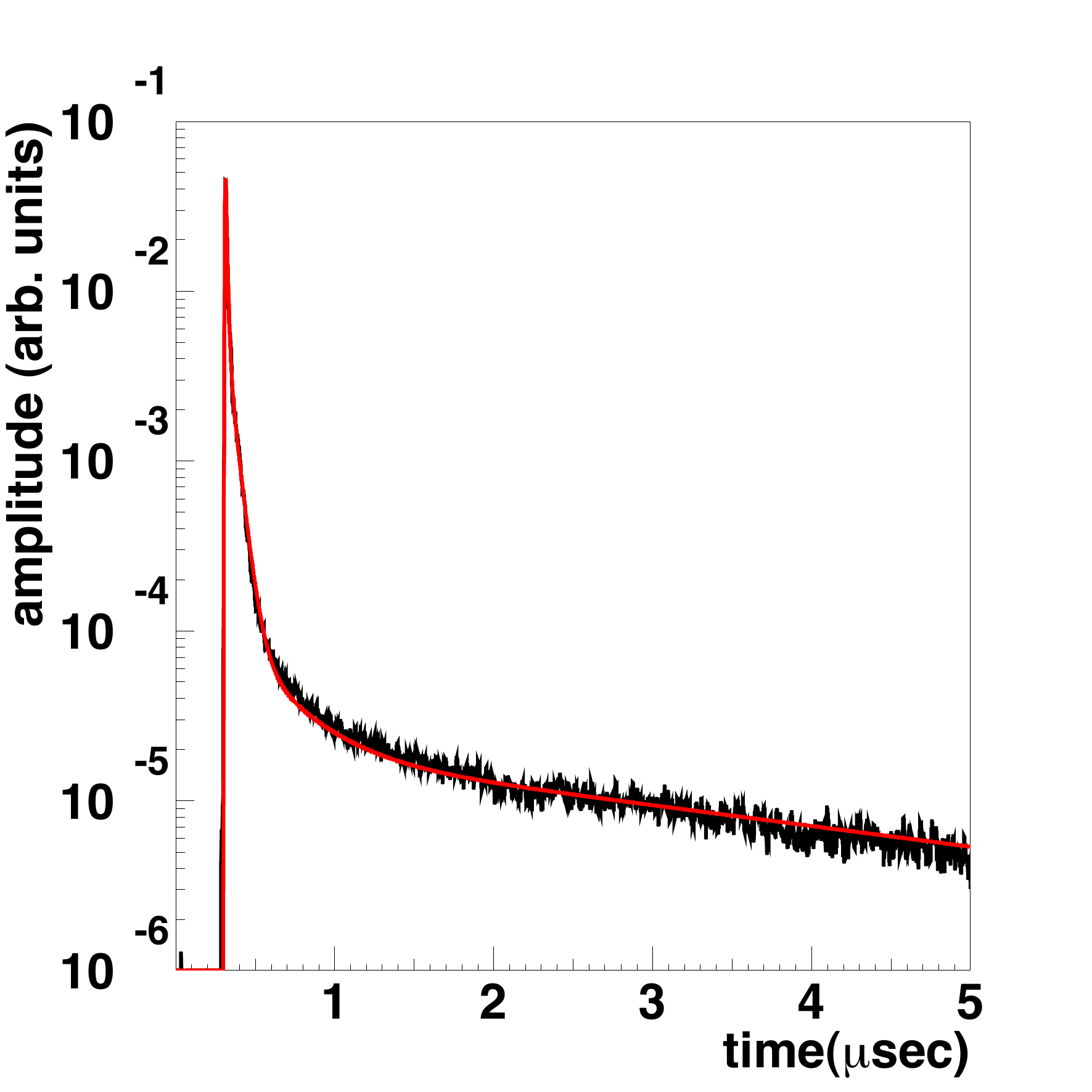}
\caption{(Color online) Response function of TPB for 127 nm photons at LAr temperature. It has been fitted with a function made of four decaying exponentials 
convoluted with a gaussian (see text). The result of the fit is represented by a red line.}
\label{fig:fit_ser}
\end{figure}

  In \cite{nitrogen} the results of a test of the effects of nitrogen contaminations in LAr  are very clearly presented.
  The detector was constituted by a PTFE cell containing about 0.7 l of  LAr lined up with a highly reflective foil (VM2000 by 3M) covered by a thin film of TPB (surface density $
  \sigma \simeq 450~\mu$gram/cm$^2$) and observed by a single 2" photomultiplier. An injection 
  system allowed to contaminate the ultra pure LAr with controlled amounts of N$_2$. The details of the experimental set-up can be found in \cite{nitrogen}.
  Contamination levels ranging from 1 ppm to 3000 ppm of N$_2$ were explored. 
    For each different contamination the LAr cell was exposed to a $\gamma$ source of $^{60}$Co. Scintillation light produced by electrons from $\gamma$ interactions was 
  wave-shifted on the surface of the cylinder and then detected by the photomultiplier. 
The average of the waveforms collected at 3000 ppm of N$_2$ contamination is shown in figure \ref{fig:fit_ser} \footnote[2]{This waveform is not shown in \cite{nitrogen}, but it 
  has been kindly granted by the authors.}.
   According to the previous discussion this waveform is obtained with a very fast 127 nm light excitation and it
    should be regarded as the time response of pure TPB to LAr scintillation photons. 

  This waveform clearly shows the expected very fast/instantaneous pulse, but also much slower components, that 
  are interpreted here as coming from the triplet-triplet interaction process in TPB. It has been clearly demonstrated that the delayed scintillation of unitary organic scintillators is not
   exponential 
  \cite{laustriat,birks3,voltz1,voltz2}, but to have a simplified quantitative idea of the time evolution of the emitted light  the waveform has been fitted with a function made of four 
  decaying exponentials, convoluted with a gaussian function that accounts for the photomultiplier response and for the electronic noise.  A more adequate treatment of the delayed 
  scintillation  will be presented in section \ref{sec:elec-alpha}.
   The result of the fit is shown in figure \ref{fig:fit_ser} with a red line and the abundance and decay time of the components are reported in table \ref{tab:fit_ser}.
 
\begin{figure*}[t]
\includegraphics*[scale=0.495]{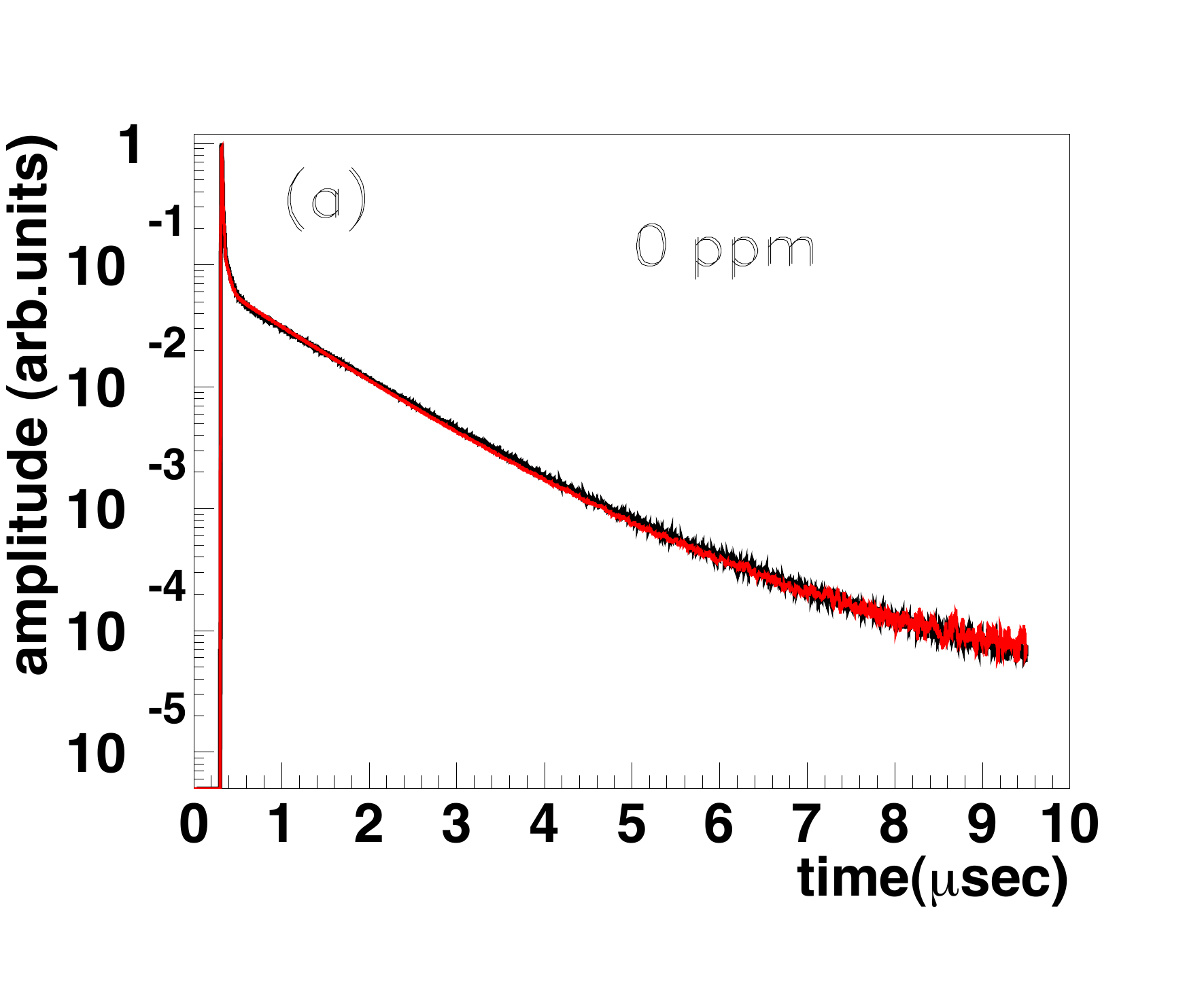}
\includegraphics*[scale=0.495]{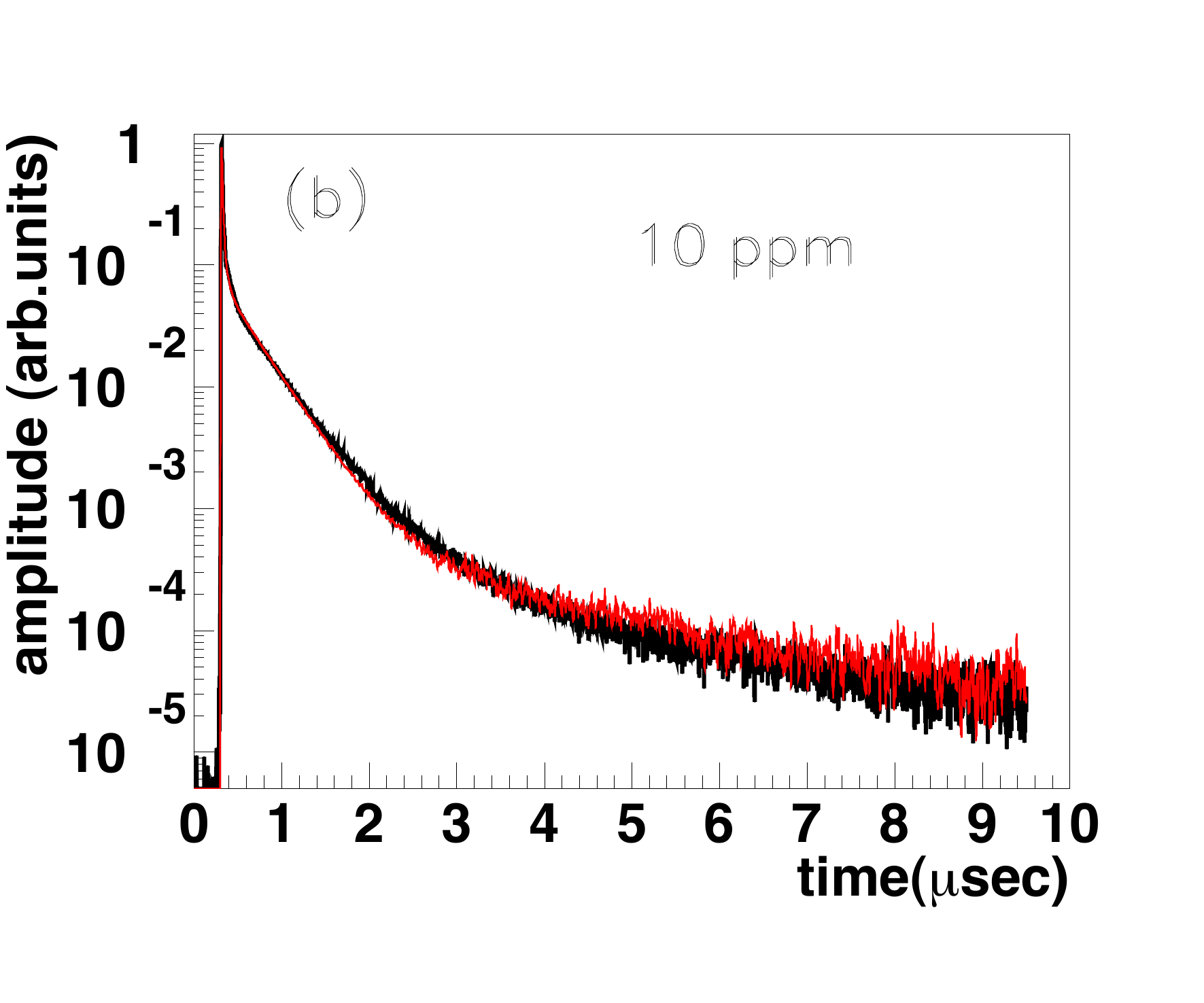}
\includegraphics*[scale=0.495]{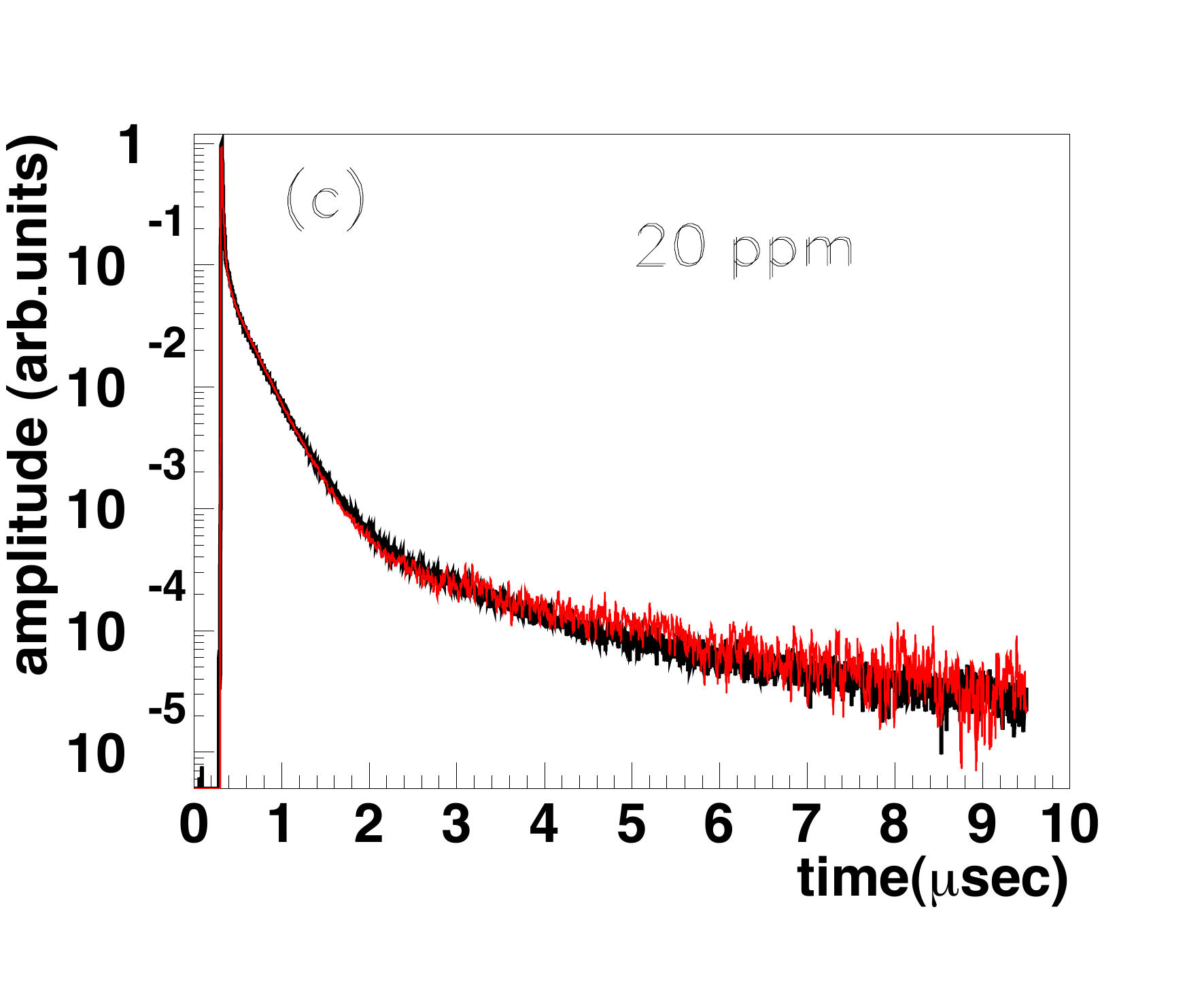}
\includegraphics*[scale=0.495]{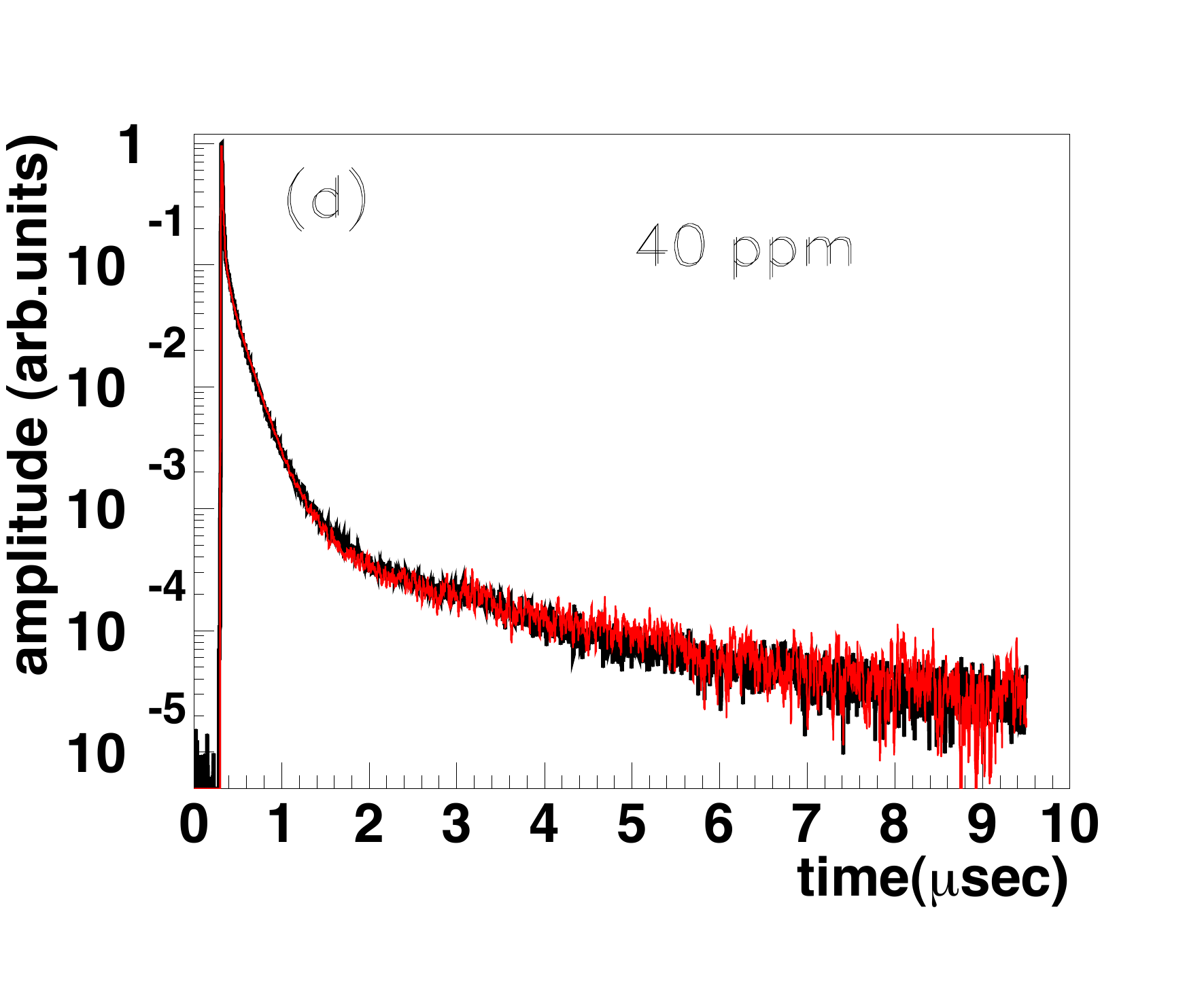}
\caption{(Color online) Average waveforms for  LAr scintillation light when excited by $^{60}$Co $\gamma$s at different levels of Nitrogen contamination: 0 ppm (a), 10 ppm (b), 20 
ppm (c), 40 ppm (d). In red the results of the fits are shown. The fit function is a  convolution of the TPB response function (3000 ppm average waveform) with the sum of two 
decaying exponentials, assumed to be the response of LAr.}
\label{fig:fit_conv}
\end{figure*}

\begin{table}[h]
\caption{Decay times and relative abundances of the components found in the decomposition into exponentials of  the response function of TPB to 127 nm photons. Only statistical 
errors from the fit are quoted.} 
\begin{ruledtabular}
\begin{tabular}{lcc}
& decay time (nsec) & abundance ($\%$)\\
\hline
Instantaneous component & 1-10 & 60 $\pm$ 1 \\
Intermediate component & 49 $\pm$ 1 & 30 $\pm$ 1 \\
Long component & 3550 $\pm500$ & 8 $\pm$ 1 \\
Spurious component & 309 $\pm$ 10 &  2 $\pm$ 1 \\
 \end{tabular}
\label{tab:fit_ser}
\end{ruledtabular}
 \end{table}  
 
 Even if unphysical, the decomposition of TPB response into exponentials allows to put in evidence some useful features. The delayed scintillation of TPB accounts for about 
 40$\%$ of the total and it is necessary to integrate the waveform for at least 140 ns to accumulate the 90$\%$ of the signal.
The slow part of the waveform fakes  essentially two exponentially decaying components, one with a 50 ns slope, the most abundant, and one with 3.5 $\mu$s.\\

  Since during the N$_2$ contamination test described in \cite{nitrogen}, data were taken for many different values of N$_2$ concentration in LAr, a very stringent test of the 
hypothesis that the waveform shown in figure \ref{fig:fit_ser} represents the response of TPB to LAr scintillation photons has been possible.
 For each level of  $[$N$_2$$]$  the average waveform shown in \cite{nitrogen}
 has been fitted with a convolution of a double exponential,  assumed to be the p.d.f. of LAr scintillation photons, with the response function of TPB. Some examples of fitted 
 waveforms are shown in figure \ref{fig:fit_conv}. In 
all the  cases the fits are almost perfect and the average waveforms are reproduced very precisely along all the time interval considered (up to 9.7 $\mu$s after the onset of the 
signal). \\
The picture that emerges from this analysis is perfectly consistent 
and there is no need of invoking exotic mechanisms of LAr scintillation, different from the two excimer states  de-excitation ($^{1}\Sigma$ and $^{3}\Sigma$), to explain all the 
features observed in the LAr scintillation waveforms.
This is true in particular for one of the points deeply analyzed in \cite{nitrogen} without reaching definite conclusions, that is the existence of an intermediate decaying component 
between the singlet and triplet  de-excitation. 

 \subsection{Test with $\beta$ and $\alpha$ interactions}
 \label{sec:elec-alpha}
 Following the idea that the waveform of figure \ref{fig:fit_ser} can represent the response of TPB to LAr scintillation photons, a dedicated experimental
 test has been performed.
 In order  to check if the observed long tail is effectively related to TPB de-excitation, a sample of  pure TPB  has been directly  irradiated with an ionizing radiation in a vacuum 
 environment. This guarantees the formation of  triplet states (through electron-ion recombination, secondary electrons excitations, $\delta$-rays...), that are in turn the precursors of 
 the delayed TPB scintillation, with a perfectly $\delta$ shaped excitation function and in a way completely independent from LAr scintillation light.\\
  
  In general the scintillation of pure organic crystals excited by an ionizing radiation can be well described with the superposition of a prompt component and of a delayed one. The 
  prompt component 
 is found to be exponentially decaying with a time constant identical to the mean lifetime $\tau_S$ of the first excited singlet state S$_1$. \\
 The time evolution of the delayed 
 component depends on the dynamics of the triplet-triplet interaction process. It can be predicted by solving the diffusion-kinetic equation for the triplet density along the 
 ionizing track, assuming a Gaussian shape, with scale parameter r$_0$, for the initial triplet distribution function \cite{laustriat,voltz1,voltz2}.  
 The asymptotic time dependence of the delayed light  (t$\gg \tau_S$) is found to be:
 \begin{equation}
 \label{eq:laustriat}
 I(t)_{delayed} \simeq \eta_S\frac{N}{[1+A~ln(1+t/t_{a})]^2 (1+t/t_a )}
 \end{equation}
where N and A are constants depending on the nature of the scintillator, $\eta_S$ is the fluorescence yield and t$_a$ is a relaxation time that is linked to the diffusion coefficient of 
 triplet states in the scintillator, D$_t$,  through the relation: t$_a$ = r$_0^2$/4D$_t$.\\
 The time evolution of the delayed light  does not depend on the particle type, but only on the dynamics of the triplet-triplet interaction process. Only the relative abundance of fast 
 and delayed components is expected to depend on the linear energy transfer, and consequently on the particle type. \\

\begin{figure}[h]
\begin{center}
\includegraphics*[scale=0.6]{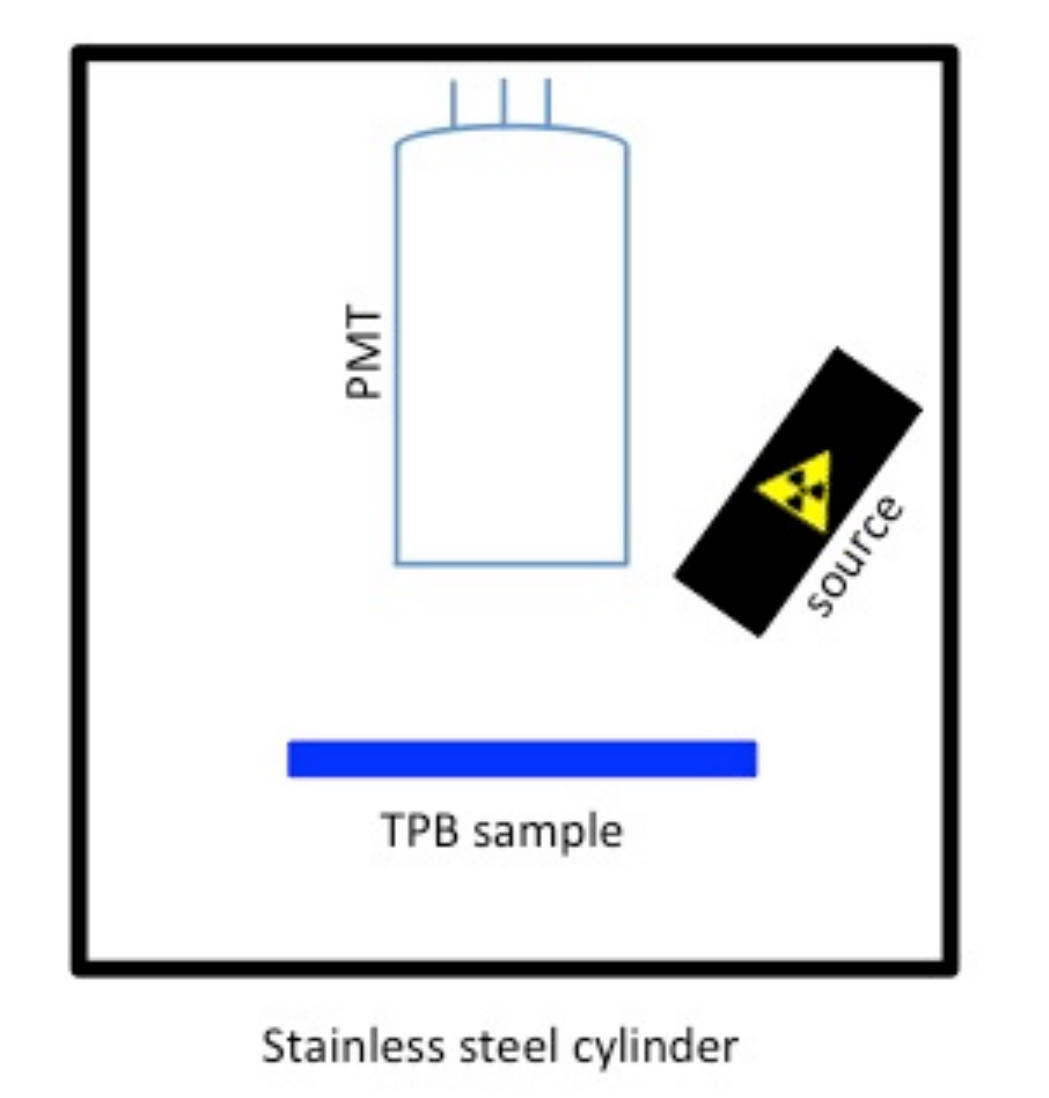}
\caption{(Color online) Scheme of the experimental set-up used to irradiate TPB films with electrons and $\alpha$ particles.}
\label{fig:water}
\end{center}
\end{figure}

 The experimental set-up that  has been built is essentially constituted by a stainless steel vacuum tight chamber that 
 hosts a 2" photomultiplier (ETL D745UA), an holder for the TPB sample and one for the radioactive source. A schematic view of the set-up is shown in figure \ref{fig:water}. The 
 TPB  sample is a film with a surface thickness $\sim$ 10$^3$ $\mu$gram/cm$^2$ evaporated on a highly reflective plastic foil (3M VM2000) circular in shape with a 
 diameter of 8 cm. The choice of having a reflective substrate below the TPB has been done to maximize the amount of light that could be collected by the photomultiplier. A 
 drawback of using 
 VM2000 is that it is weakly emitting light when irradiated by ionizing particles. The film has been produced at LNGS with a dedicated evaporation system. More details can be
  found in \cite{aging}. The distance between the sample  and the photomultiplier is 5 cm. Two sources have been alternatively used:
 \begin{itemize}
 \item an $\alpha$ source made of an alloy of Uranium and Aluminum that emits $\alpha$ particles with a continuous spectrum with end point around 5 MeV;
 \item a $^{90}$Sr $\beta ^-$ source with a Q value of 546 keV.
 \end{itemize}

During each measurement the stainless steel chamber has been evacuated down to a pressure of 5$\times$10$^{-5}$ mbar to allow $\alpha$ particles and electrons to 
 hit the TPB film without being captured by air. Scintillation signals detected by the photomultiplier have been sent to a fast Waveform Recorder (Acqiris, DP235 Dual-Channel PCI 
 Digitizer Card). The signal waveforms passing a threshold set at a level corresponding to few photo-electrons have been recorded with sampling time of 1 ns over a full record 
 length of 10 $\mu$s.\\ 
 All the measurements have been performed at room temperature, since it has been shown in \cite{veloce}  that the time dependence of the late components of 
 TPB fluorescence excited by $\alpha$ particles does not change appreciably at LAr temperature.
 
 \subsection{Data analysis and comparison}
\begin{figure}[h]
\includegraphics*[scale=0.52]{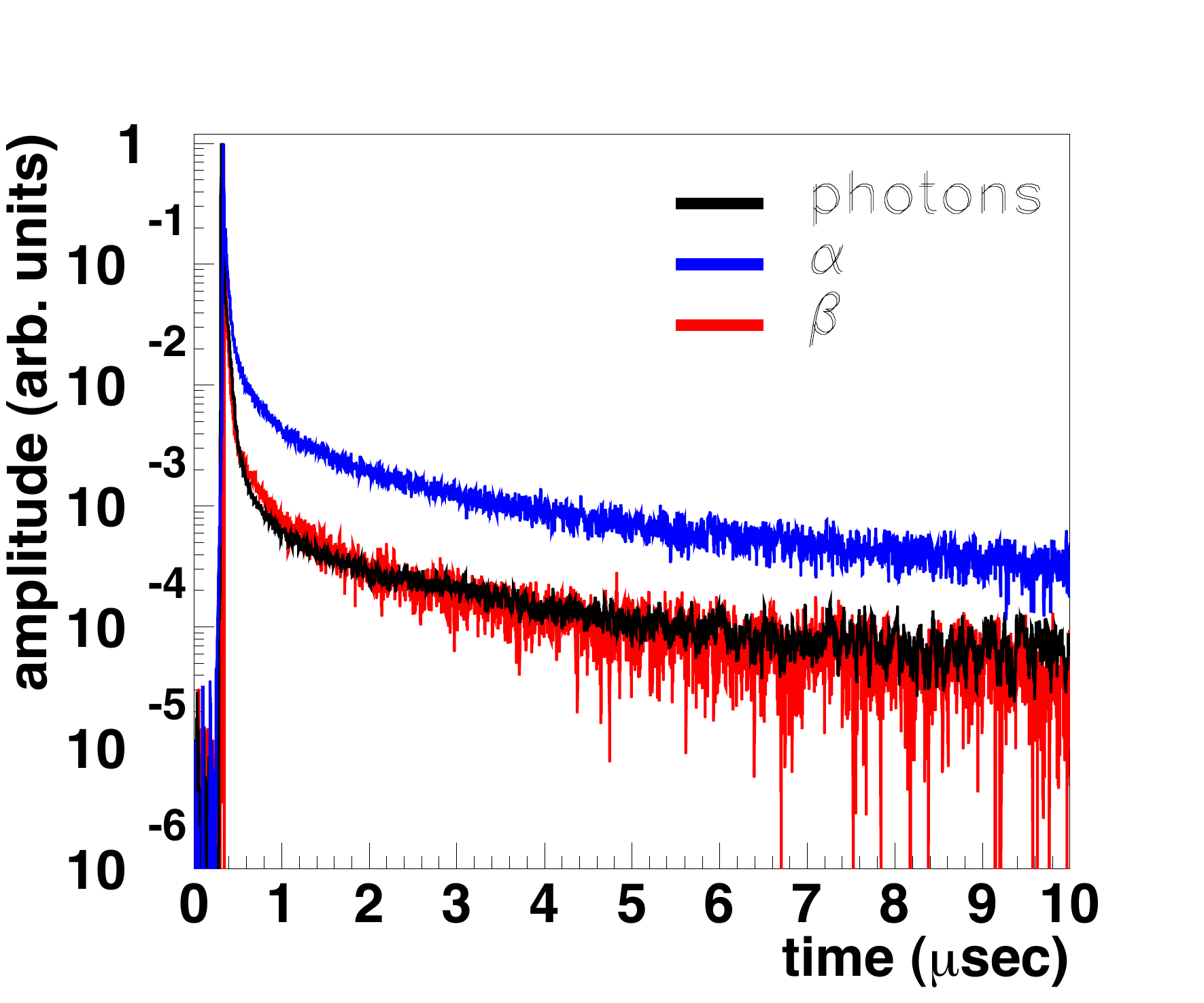}
\caption{(Color online) Red line: average waveform obtained by irradiating a TPB film with electrons. Blu line: average waveform 
obtained by irradiating a TPB film with $\alpha$ particles. Black line: average waveform obtained irradiating TPB with LAr scintillation light quenched by 3000 ppm of Nitrogen.}
\label{fig:gamma-e-alpha}
\end{figure}

The average waveforms for the $\beta$ and $\alpha$ particles tests have been calculated applying simple cuts to eliminate waveforms that present saturations, afterpulses or 
multiple signals and the result is shown in figure \ref{fig:gamma-e-alpha}.\\
The red curve refers to $\beta$ irradiation and the blue one to $\alpha$, while the black curve is the one obtained with LAr scintillation light, already shown in figure \ref{fig:fit_ser}, 
reported here for visual comparison. Beta and VUV photons curves are nicely overlapped as is observed also for sodium salicylate \cite{baker}.
The small differences found around 300-400 ns after the onset of the signal could be ascribed to a small effect of fluorescence of the plastic substrate of the TPB film, since the
electrons have enough energy to reach and traverse it. 
$\alpha$ particles, instead, show a much higher abundance of  delayed component.\\
 In order to investigate the details of the tails of the three curves and to check if  they are compatible with each other a {\it single photon counting like procedure} has been adopted. 
 The classical {\it Coincidence Single Photo-electron Counting technique} has been used many  times in scintillation lifetime measurements \cite{himi,hitachi,carvalho}.     
 The recorded waveforms  allow to implement an offline version of this technique. Starting at 170 ns after the onset of a triggered signal, a single photo-electron finding algorithm is 
 run through the waveform and for each photo-electron pulse (defined by appropriate cuts) the arrival time is stored. In order to minimize the pile-up of single photo-electrons, that is 
 a time dependent  effect, only waveforms with a total integral below 40  photo-electrons in the case of electron and photon excitations and 20 in the case of $\alpha$s are 
 considered. In this way the pile up probability in the first 30 ns results to be below 5$\%$ in all three cases.
 
\begin{figure}[h]
\includegraphics*[scale=0.52]{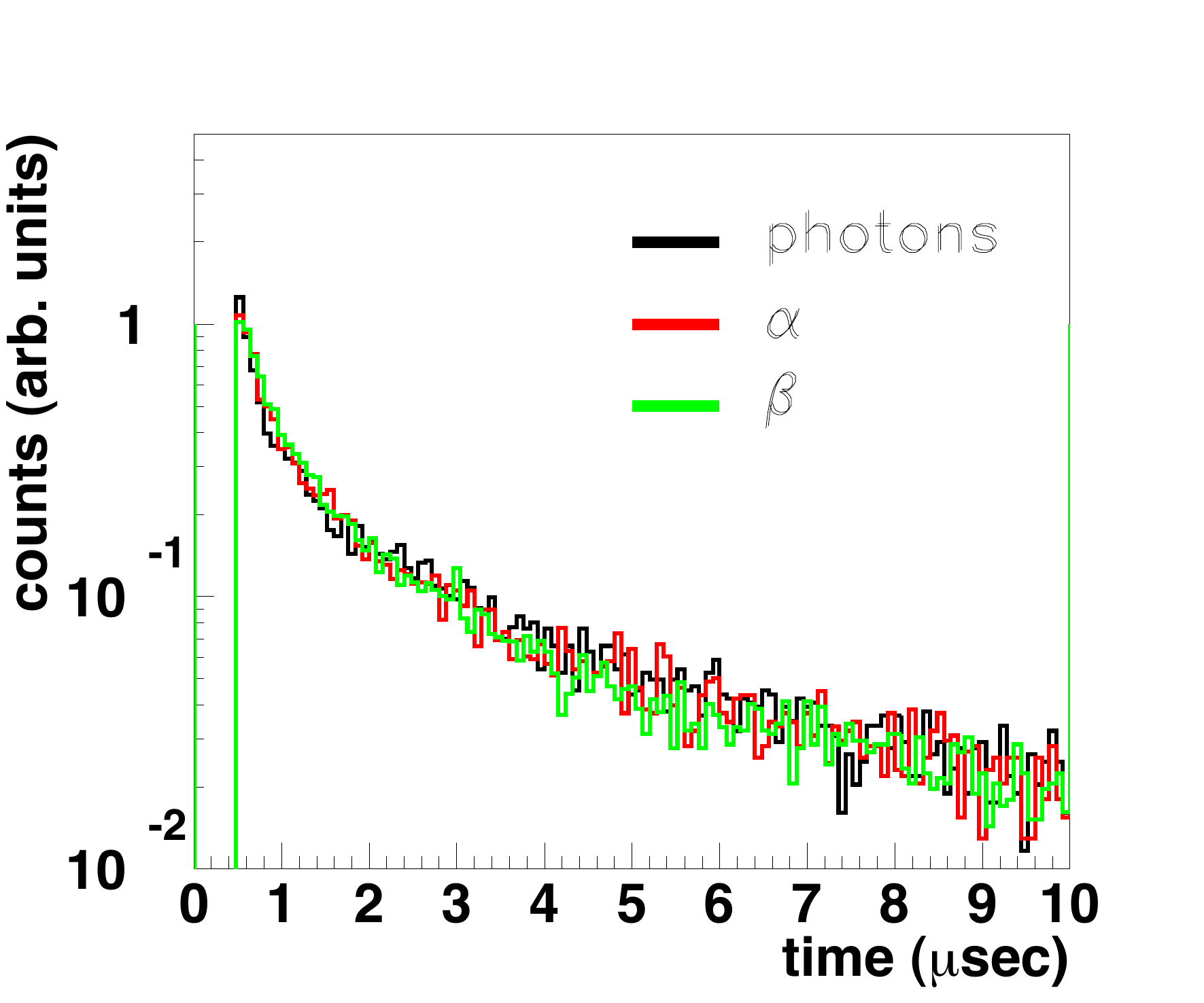}
\caption{(Color online) Normalized histograms of the arrival time of single photons for photon (black curve), $\alpha$ (red curve) and electron (green curve) excitations of TPB.}
\label{fig:single}
\end{figure}
\begin{figure}[h]
\includegraphics*[scale=0.52]{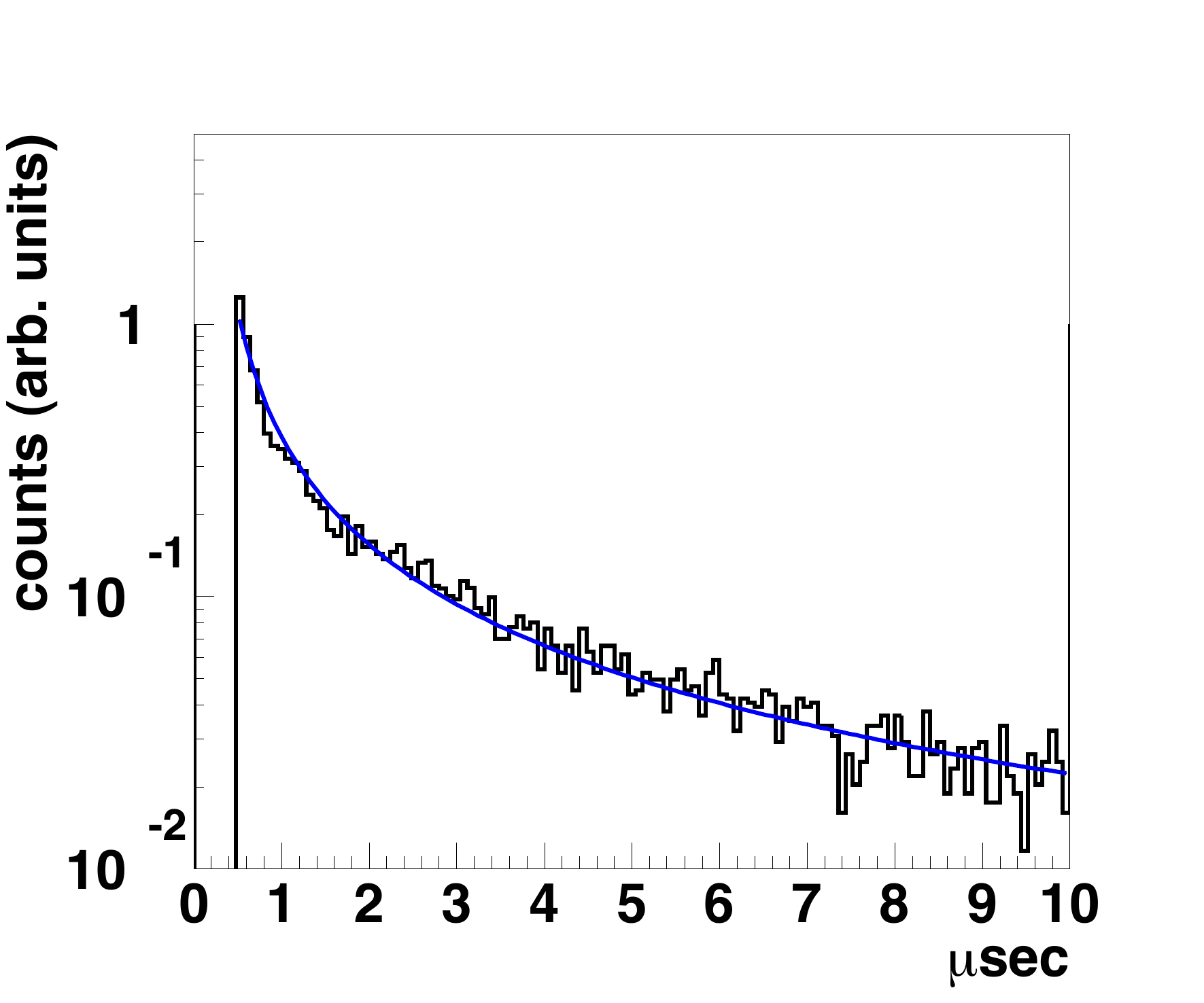}
\caption{(Color online) Histogram of arrival times of single photons for TPB excited by  VUV photons. The blue line represents the result  of the fit  with the function of equation \ref{eq:laustriat}.}  
\label{fig:single_fit}
\end{figure}

 
 The time of arrival of photo-electrons have been accumulated in three histograms, that are shown in figure \ref{fig:single}. They are  almost perfectly overlapped. The histogram 
 related to VUV photons has also been fitted with the function of  equation  \ref{eq:laustriat} where A, t$_a$ and the product $\eta_S$N are left as free parameters. The result of the
  fit is shown in figure \ref{fig:single_fit} with a blue line. The best fit  values for A and t$_a$ are 0.22 and 51 ns respectively,  not so different from what is found for anthracene 
  \cite{bollinger} - A=0.25 and t$_a$=40 ns - and for stilbene  \cite{wasson} - A=0.25 and t$_a$=80 ns.\\
 
  The perfect compatibility of the delayed scintillation of TPB excited by LAr scintillation photons  with that induced by ionizing particles clearly demonstrates that it is generated by a 
  triplet-triplet interaction mechanism, started by the ionization of TPB by 127 nm photons. The TPB response function to LAr scintillation light is not a fast decaying 
  exponential, as it is found for near UV excitation, but has a much more complex structure with a delayed component that has a non-exponential shape and that accounts for about 
  the 40$\%$ of the emitted radiation.

\section{Discussion}
The waveform shown in figure \ref{fig:fit_ser} confidently represents the response function in time of TPB to 127 nm. Having demonstrated that convoluting it with the sum of only 
two exponentially decaying functions allows to reproduce the average waveforms measured at any level of contamination is a remarkable point.
It demonstrates that the time evolution of LAr scintillation light can be described as the sum of only two decaying exponentials, originated from  the de-excitation of the lowest lying 
triplet and singlet states of the Ar$^*_2$ excimer. The observation of an intermediate component with a decay slope in the range of  50-100 ns often reported in literature 
 \cite{nitrogen,lippincott,amsler} can be totally ascribed to the fluorescence of TPB. It is the slow TPB de-excitation following the fast LAr scintillation pulse.
  It  has been shown in section \ref{sec:vuv_lar_photons}, in fact, that the TPB  response function fakes a 50 ns component, if one attempts to decompose it into exponentials. Also 
  the difficulty of determining unambiguously the slope of the slow scintillation component of LAr can be a consequence of the use of TPB or of wavelength shifters in general. The 
  long tail in the TPB response function, resembling a 3.5 $\mu$s exponential, distorts the slow component of LAr scintillation photons and consequently any technique to measure 
  its decay constant brings inside a certain amount of uncontrolled systematics  if the  effect of TPB is not properly deconvolved. It has been shown in \cite{nitrogen}, for example, 
  that the use of two slightly different fitting procedures lead to two quite different values of the LAr slow decay slope. In this respect the most reliable value  appears to be the one of  
  1300 $\pm$ 60 ns reported in \cite{heindl}, measured without shifter, and in \cite{nitrogen} with sophisticated deconvolution techniques.\\
 
LAr is used in several experiments for the direct Dark Matter detection mainly because it allows to reject efficiently $\gamma$ and $\alpha$ background with respect to nuclear 
recoil events that could be due to a WIMP (Weakly Interactive Massive Particle) signal \cite{warp_100, darkside, ardm, deap} . In fact different ionizing particles produce very 
different scintillation signals in LAr. In particular the relative abundance of the fast  to  slow scintillation  components are different for electrons, $\alpha$s and nuclear recoils, being 
respectively 1/3, 1.3 and 3 \cite{hitachi, lippincott, phd_rob}.
The most  widely used technique to exploit the pulse shape discrimination of LAr is based on the calculation of the prompt fraction of light in the signals. A factor, usually 
called F$_{prompt}$, is defined as follows:
\begin{equation}
F_{prompt}=\frac{\int_0^{t^*}I(t) dt }{\int_0^{\infty}I(t) dt }
\end{equation} 

where I(t) is the intensity of the detected scintillation signal measured in photo-electrons and {\emph t$^* $} is the integration time of the prompt signal that maximizes the 
separation among different particles . It has been found experimentally by many groups that the optimum value of  {\emph t$^* $} is around 100 ns \cite{warp_100, darkside, lippincott, phd_rob}. This an indirect but clear confirmation of the existence of the delayed fluorescence of TPB.\\
According to the picture that emerges from this work, the scintillation of LAr shifted by TPB can be described by the p.d.f. :
\begin{equation}
\label{eq:TPB}
L(t) = A\times S(t) + (1-A)\times T(t)
\end{equation}   

where {\emph S(t)} and {\emph T(t)} are the fast and slow exponential components of LAr both convoluted with the TPB response and A is the fraction of prompt light. It is 
straightforward to prove that {\emph t$^* $} can be found by solving the equation:
\begin{equation}
\label{eq:t_star}
S(t^*) = T(t^*)
\end{equation}

\begin{figure}[h]
\includegraphics*[scale=0.52]{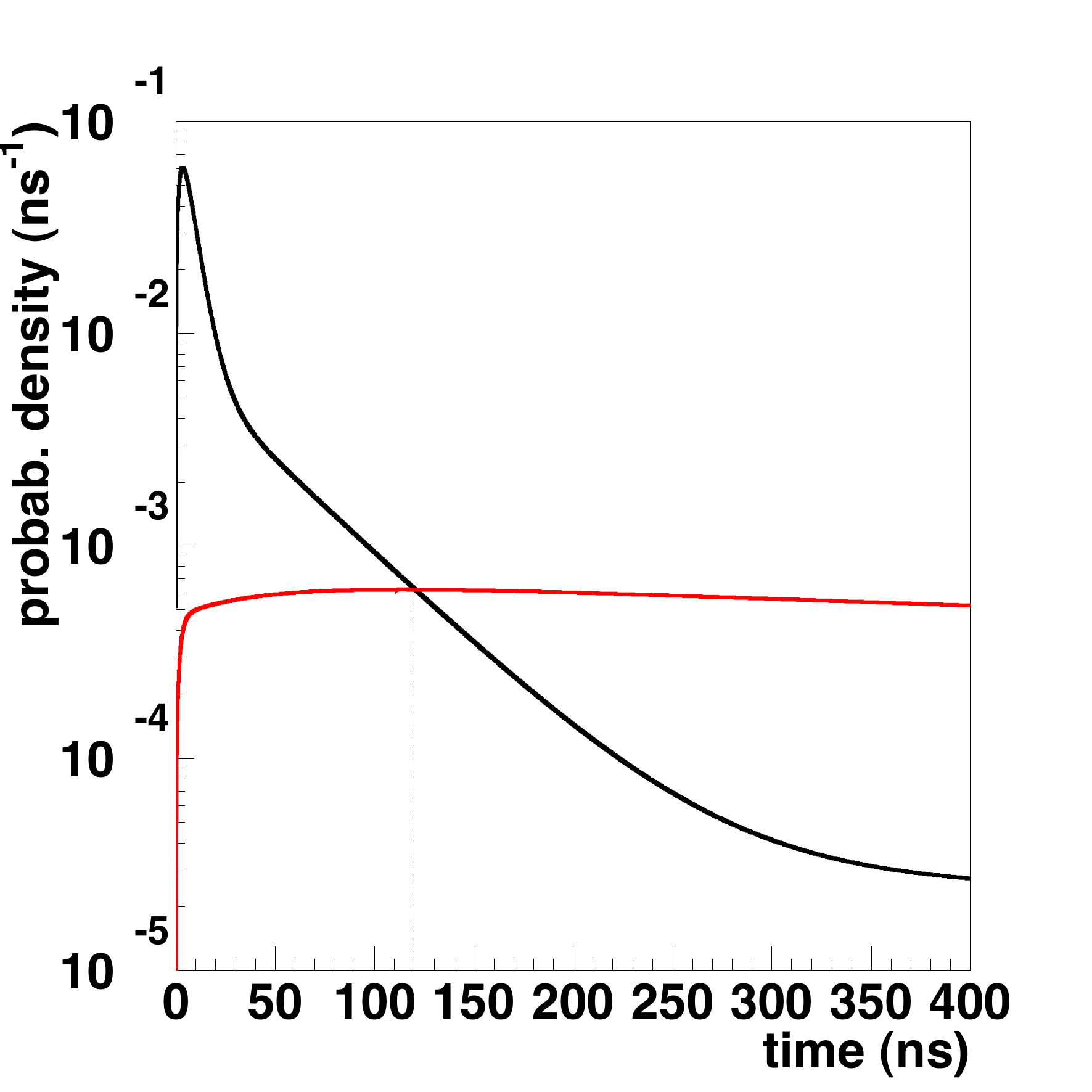}
\caption{(Color online) Graphical solution of equation \ref{eq:t_star}. In black the function {\emph S(t)}  and in red {\emph T(t)}. The solution is represented by the crossing point of 
the two functions, that is found around 120 ns.}  
\label{fig:f_prompt}
\end{figure}

The graphical solution of equation \ref{eq:t_star} is shown in figure \ref{fig:f_prompt} and leads to a value of  {\emph t$^* $} around 120 ns, perfectly compatible with the 
experimental observations\footnote[3]{in this case, for simplicity, it has been used the representation of the TPB response function in terms of exponentials. }. 

If the delayed scintillation of TPB were not present, {\emph S(t)} and {\emph T(t)} could be described by two exponentials, with characteristic times of
$\tau_S$ $\sim$ 6 ns and $\tau_T$ $\sim$ 1300 ns, that is:
\begin{equation}
\label{eq:no_TPB}
L(t) = A\times \frac{1}{\tau_S}e^{{-\frac{t}{\tau_S}}}+(1-A)\times \frac{1}{\tau_T}e^{{-\frac{t}{\tau_T}}}
\end{equation}

 In this case equation \ref{eq:t_star} could be analytically solved and, in the limit of $\tau_T \gg \tau_S$, one obtains:
\begin{equation}
t^*=\tau_S~ln\frac{\tau_T}{\tau_S} 
\end{equation}

that returns a value for  {\emph t$^* $} around 32 ns, three times lower than its experimental value.\\
The delayed fluorescence of TPB has also the effect of deteriorating the discrimination capability of LAr that could be  obtained in the ideal case of a direct detection of the VUV 
photons. This is because a fraction of the prompt light is delayed and the two LAr scintillation components are more mixed. A rough calculation can be very explicative.
Without the shifter,  the average value of F$_{prompt}$ for electrons and neutrons can be easily calculated. Assuming a fraction {\emph A} for the prompt scintillation of 
electrons and neutrons of 0.25 and 0.75 respectively and considering a value of {\emph t$^* $} of 32 ns, simple exponential integrations of equation \ref{eq:no_TPB} leads to
 F$_{prompt}$ values of 0.27 and  0.75 with a  difference $\Delta^{pure}$ = 0.48. In the usual situation, that is with TPB, a numerical integration of the p.d.f. of  equation \ref{eq:TPB} 
 up to  {\emph t$^* $} = 110 ns leads to F$_{prompt}$ values of 0.27 and 0.67 for electrons and neutrons  respectively with a difference of $\Delta^{TPB}$ = 0.4. The use of the 
 shifter worsens the separation between electrons and neutrons of about 17$\%$.\\

 TPB is widely used also to downshift the scintillation photons of liquid Helium (LHe) and Neon (LNe) \cite{lippi_neon, mc_he}, that have energies higher than that of LAr ones. This 
 suggests that the same mechanism of TPB delayed fluorescence should be active also in these cases. Despite the fact that it was never explicitly noticed, it could be useful in
 explaining some of the not fully clarified features of LNe and LHe scintillation. It has been shown in \cite{mc_he} that the scintillation of LHe 
 has a non-trivial time structure. In addition to the expected fast and slow components originated by the 
 de-excitation of the lowest lying triplet and singlet states of the excimer He$_2^*$, with decay times of  $\sim$ 10 s and $\sim$ 10 ns respectively, two more components are 
 observed, one exponential with a characteristic time of 1.6 $\mu$s and one non exponential that decays as $t^{-1}$. The delayed scintillation of TPB could represent a non 
 negligible contribution to these scintillation components since it is active exactly in the same time range. Even if  a direct and quantitative comparison is not reasonable, due to 
 possible effects related to the large difference in temperature, the $t^{-1}$ component resembles the asymptotic behavior of the TPB response function measured in LAr.
On the other side the experimental evidence that the 1.6 $\mu$s exponential decay 
is  different for cold Helium gas than for LHe  \cite{mc_he} demonstrates that some additional process must be active inside the LHe and the observed features can not be 
completely explained  by the delayed TPB fluorescence.\\
 
 A similar situation is found for LNe scintillation, where two approximately exponential intermediate components between the singlet and triplet Ne$_2^*$ de-excitations are found  
 \cite{lippi_neon}, with characteristic times in the range of 100 ns and 1 $\mu$s. Their origin is not clear, but it is plausible that TPB delayed scintillation can contribute to explain at 
 least a fraction of it. 
 
\section{Conclusions}
 This work shows the experimental evidence of the existence of a delayed scintillation component of TPB when excited by the VUV radiation of LAr. The production of 
  the triplet states, that are the precursors of the delayed light, is made possible by the high energy of LAr scintillation photons that can ionize the organic molecules of TPB. Its time 
  dependence has been measured with an experimental set-up that uses LAr scintillation light quenched by nitrogen contaminations to excite TPB. It has been compared to the time 
  behavior  of the delayed light of TPB when excited by $\beta$ and $\alpha$ particles and they have been found to be perfectly compatible among each other. The time shape
  of the light emission has also been found to be  consistent with what expected from the delayed luminescence of a unitary scintillator as described in literature.\\
  
  This experimental fact sheds some light on the most relevant incongruities that have been reported in the past years concerning the time dependence of LAr scintillation light. 
  Namely the presence of an intermediate component with a decay time in the range of 50-100 ns and the ambiguity in the determination of the decay time of the slow 
  component, for which values ranging from 800 ns to 1600 ns have been reported.\\ 
   
  LAr scintillation is often used for particle discrimination since the relative abundance of the fast and slow components strongly depends on the particle type. The use of TPB tends 
  to worsen this feature of LAr since a consistent part of the prompt light is delayed and the two populations are more mixed.\\ 
  TPB is an exceptionally efficient shifter for the VUV scintillation light of LAr  and also a convenient one for its emission wavelength around 430 nm matching the quantum 
  efficiency of many standard photomultipliers, but it has some drawbacks when the time features of the scintillation signals are used since they result to be slightly distorted.

  \section{acknowledgments}{The author acknowledges Prof. F. Cavanna for his contribution to this work with discussions and ideas; Prof. R. Francini for the discussions on 
 molecular processes; Dr. A. A. Machado for her help and patience and  Dr. N. Canci for his contribution to the measurements with $\alpha$ and $\beta$ particles.\\
 This paper is dedicated to the memory of Antonio Di Filippo.}


\begin{thebibliography}{99}

\bibitem{warp_100}  {WArP Coll.}, \emph{The WArP experiment}, {\emph{Journal of Physics: Conference Series}  {\bf 203} (2010) 012006.} 

\bibitem{icarus} {ICARUS Coll.}, \emph{Underground operation of the ICARUS T600 LAr-TPC: first results}, {\emph{JINST} {\bf 6} (2011) P07011}       

\bibitem{darkside} {DarkSide Coll.}, \emph{First Results from the DarkSide-50 Dark Matter Experiment at Laboratori Nazionali del Gran Sasso}, {arXiv:1410.0653}

\bibitem{micro}{MicroBooNE Coll.}, \emph{A Proposal for a New Experiment Using the Booster and NuMI Neutrino Beamlines: MicroBooNE}, {http://www-microboone.fnal.gov/public/MicroBooNE\_10152007.pdf}

\bibitem{phd_rob}{Acciarri R.}, \emph{Measurement of the scintillation time spectra and Pulse Shape Discrimination of low-energy electron and nuclear recoils in liquid Argon with the WArP 2.3 lt detector}, {\emph L'Aquila University}, (2010) PhD thesis.

\bibitem{tpb_paper} {Francini R. et al.}, \emph{VUV-Vis optical characterization of Tetraphenyl-butadiene films on glass and specular reflector substrates from room to liquid Argon temperature} , {\emph {JINST} {\bf 8} (2013) P09006} 

\bibitem{lally} {Lally C.H. et al.}, \emph{UV quantum efficiencies of organic fluors}, {\emph {Nucl. Instr. and Meth. in Phys. Res. B } {\bf 117} (1996) 421}

\bibitem{mckinsey} {McKinsey D.N. et al}, \emph{Fluorescence Efficiencies of Thin Scintillating Films in the Extreme Ultraviolet Spectral Region}, {\emph {Nucl. Instr. and Meth. in Phys. Res. B } {\bf 132} (1997) 351}

\bibitem{gehman} {Gehman V.M et al}, \emph{Fluorescence Efficiency and Visible Re-emission Spectrum of Tetraphenyl Butadiene Films at Extreme Ultraviolet Wavelengths}, {\emph {Nucl. Instr. and Meth. in Phys. Res. A } {\bf 654} (2011) 116}

\bibitem{flournoy} {Flournoy J.M. et al}, \emph{Substituted tetraphenylbutadienes as fast scintillator solutes}, {\emph  {Nucl. Instr. and Meth. in Phys. Res. A } {\bf 351} (1994) 349}

\bibitem{camposeo} {Camposeo A. et al.}, \emph{Random lasing in an organic light-emitting crystal and its interplay with vertical cavity feedback}, {\emph {Laser Photonics Rev.} {\bf 8}  (2014) No. 5, 785}
 
\bibitem{laustriat} {Laustriat G.},  \emph{The luminescence decay of organic scintilator}, {\emph  {Molecular Crystal } {\bf 4} (1968) 127.}

\bibitem{birks} {Birks J.B.}, \emph{The theory and practice of scintillation counting}, {\emph  {Pergamon Press, Oxford} (1964).}

\bibitem{birks2} {Birks J.B.}, \emph{Photophysics of Aromatic Molecules}, {\emph  {Wiley-Interscience, London and New York} (1970).}

 \bibitem{kafer} {Kafer D.},   \emph{Characterization and Optimization of Growth and Electronic Structure of Organic Thin Films for Applications in Organic Electronics}, {\emph Ruhr-University Bochum}, (2008) PhD thesis (available at http://www-brs.ub.ruhr-uni-bochum.de/netahtml/HSS/Diss/KaeferDaniel/diss.pdf)
 
 \bibitem{hill} {Hill I.G. et al.}, \emph{Charge-separation energy in films of p-conjugated organic molecules}, {\emph {Chemical Physics Letters} {\bf 327}  (2000) 181.}

\bibitem{baker} {Baker G.J. et al}, \emph{Time dependence of sodium salicylate luminescence excited by vuv photons, x-rays and $\beta$ particles: magnetic field effects}, {\emph {J. Phys. B: At. Mol. Phys.} {\bf 20}  (1987) 305.}

\bibitem{klein} {Klein G. and Carvalho M.J.}, \emph{Highly excited states devcay in p-terphenyl crystals, magnetic field effect investigation}, {\emph {Chem. Phys. Lett.} {\bf 51} , No. 3 (1977) 409.}

\bibitem{nitrogen} {Acciarri R. et al.},  \emph{Effects of Nitrogen contamination in liquid Argon}, {\emph{JINST} {\bf 5} (2010) P06003} 

\bibitem{birks3}  {Birks J.B.},  \emph{Liquid scintillator solvents}, {\emph  {Proceedings of the International Conference of Organic Scintillators and Liquid Scintillation Counting}, 
Academic Press, New York (1971).} 


\bibitem{voltz1} {Voltz R. and Laustriat G.}, \emph{Radioluminescence des milieux organiques I. ƒtude cinetique}, {\emph {J. Phys. France} {\bf 29} (1968) 159.}

\bibitem{voltz2} {Voltz R. and Laustriat G.}, \emph{Radioluminescence des milieux organiques II. Verification experimentale de l'etude cinetique}, {\emph {J. Phys. France} {\bf 29} (1968) 297.}

\bibitem{lippincott} {Lippincott W.H. et al}, \emph{Scintillation time dependence and pulse shape discrimination in liquid argon}, {\emph{Phys. Rev. C} {\bf 78} (2008) 03580.} 

\bibitem{amsler} {Amsler C. et al.}, \emph{Luminescence quenching of the triplet excimer state by air traces in gaseous argon}, {\emph{JINST} {\bf 3} (2008) P02001} 

\bibitem{aging}  {Acciarri R. et al.}, \emph{Aging studies on thin tetra-phenyl butadiene films}, {\emph{JINST} {\bf 8} (2013) P10002} 

\bibitem{heindl} {Heindl T. et al.},  \emph{The scintillation of liquid argon}, {\emph{EPL} {\bf 91} (2010) 62002.}

\bibitem{morikawa} {Morikawa E. et al.},   \emph{Argon, Krypton an Xenon excimer luminescence: from dilute gas to the condensed phase}, {\emph{J. Phys. Chem.} {\bf 91} (1989) 1469.} 
 
 \bibitem{kubota} {Kubota S. et al.} ,  \emph{Recombination luminescence in liquid Ar and Xe}, {\emph{Phys. Rev. B} {\bf 17} (1978) 2762.} 
 
 \bibitem{doke} {Doke T.}, \emph{Fundamental properties of liquid Argon, Krypton and Xenon as Radiation detector media}, {\emph{Portgal. Phys.} {\bf 12} (1981) 9.} 
 
 \bibitem{himi} {Himi S. et al.},  \emph{Liquid and solid Argon, and Nitrogen doped liquid and solid Argon scintillators}, {\emph  {Nucl. Instr. and Meth. in Phys. Res.} {\bf 203} (1982) 153.}
 
 \bibitem{veloce} {Veloce L.M.}, \emph{An Investigation of Backgrounds in the DEAP-3600 Dark Matter Direct Detection Experiment}, {\emph QueenÕs University - Kingston, Ontario, 
 Canada}, (2013) MD thesis.
 
  \bibitem{hitachi} {Hitachi A. et al.}, \emph{Effect of ionization density on the time dependence of luminescence from liquid argon and xenon}, {\emph  {Phys. Rev. B} {\bf 27} (1983)  5279.}
 
 \bibitem{carvalho} {Carvalho M.J. et al},  \emph{Luminescence decay in condensed Argon under high energy excitation}, {\emph  {J. Lumin} {\bf 18-19 } (1979)  487.}
 
 \bibitem{bollinger} {Bollinger L.M. and Thomas G.E.},  \emph{Measurement of the Time Dependence of Scintillation Intensity by a Delayed Coincidence Method}, {\emph  {Rev. Sci. Instrum.} {\bf 32} (1961) 1044.}
 
 \bibitem{wasson} {Wasson M.M.and Memo M.},   {\emph  {Aere, Harwell}  (1962) 1153.}
  
 \bibitem{ardm}  {Badertscher A. et al.},  \emph{Status of the ArDM Experiment: First results from gaseous argon operation in deep underground environment}, arXiv:1307.0117 
 
 \bibitem{deap} {Boulay M.G.},  \emph{DEAP-3600 Dark Matter Search at SNOLAB}, {\emph  {J. Phys. Conference Series} {\bf 375} (2012) 012027.}
 
 \bibitem{lippi_neon} {Lippincott W. H. et al.}, \emph{Scintillation yield and time dependence from electronic and nuclear recoils in liquid neon}, {\emph {Phys. Rev. C} {\bf 86} (2012) 015807.}
 
 \bibitem{mc_he}{McKinsey D. N. et al.}, \emph{Time dependence of liquid-helium fluorescence}, {\emph {Phys. Rev. A} {\bf 67} (2003) 062716.}



  
\end{thebibliography}
\end{document}